# $MoS_2$-graphene in-plane contact for high interfacial thermal conduction

Xiangjun Liu, Junfeng Gao, Gang Zhang,* Yong-Wei Zhang

Institute of High Performance Computing, A*STAR, Singapore, 138632

**ABSTRACT**

Recent studies showed that the in-plane and inter-plane thermal conductivities of two-dimensional (2D) $MoS_2$ are low, posing a significant challenge in heat management in $MoS_2$-based electronic devices. To address this challenge, we design the interfaces between $MoS_2$ and graphene by fully utilizing graphene, a 2D material with an ultra-high thermal conduction. We first perform *ab initio* atomistic simulations to understand the bonding nature and structure stability of the interfaces. Our results show that the designed interfaces, which are found to be connected together by strong covalent bonds between Mo and C atoms, are energetically stable. We then perform molecular dynamics simulations to investigate the interfacial thermal conductance. It is found surprisingly that the interface thermal conductance is high, comparable to that of graphene-metal covalent-bonded interfaces. Importantly, each interfacial Mo-C bond serves as an independent thermal channel, enabling the modulation of interfacial thermal conductance by controlling Mo vacancy concentration at the interface. The present work provides a viable route for heat management in $MoS_2$ based electronic devices.

# INTRODUCTION

Two-dimensional (2D) materials and their heterostructures have attracted a great deal of research interest recently. It is highly likely that this trend will continue in many research communities, such as in physics, chemistry, and materials science.[1,2] Monolayer $MoS_2$, a member of 2D materials family, is under the spotlight in recent years. As a semiconductor with a direct bandgap (1.8 eV), monolayer $MoS_2$ is promising in electronic and photonic device applications, including transistors, light-emitters, photovoltaic and photodetectors.[3] With successful growth of $MoS_2$ on insulating substrates,[4] and significant improvement in its mobility at both low temperature and room temperature,[5–8] it is expected that high-performance field-effect transistors based on monolayer $MoS_2$ will be realized in the near future. In $MoS_2$-based integrated devices, naturally, their thermal management will become vitally important. On one hand, the highly localized Joule heating in the ultrathin channels with atomic thickness can easily create "hot spots". On the other hand, previous experimental[9–11] and theoretical studies[12–16] revealed that the thermal conductivity of monolayer $MoS_2$ is very low. This poses a significant challenge for efficient thermal management of $MoS_2$-based integrated devices.

Graphene, another member in the 2D material family, possesses an ultra-high thermal conductivity.[17–20] A probable solution is to construct a $MoS_2$-graphene heterostructure by fully utilizing the ultra-high thermal conducting graphene to speed up the heat dissipation of "hot-spot" in $MoS_2$ devices. Then, several important questions arise: What is the nature of bonding between $MoS_2$ and graphene? Is the interface structure energetically stable? What is the thermal conductance across such interfaces? And what are the effects of temperature and interfacial defects on the interfacial thermal conductance? Clearly, answers to these questions are not only of significant scientific interest in understanding the structural and thermal properties of $MoS_2$-graphene in-plane heterostructure, but also of great impact on addressing the thermal management issues in $MoS_2$ integrated devices.

In this work, using first-principles calculations, we systematically explore the structures and energetics of $MoS_2$-graphene interfaces. We find that there are strong covalent bonds formed between carbon and Mo atoms. As a result, the stability of the interfaces is high, comparable with many well-observed heterostructures. Moreover, using molecular dynamics (MD) simulations, we further study the thermal conduction across the interfaces between monolayer

MoS$_2$ and graphene. Interestingly, we find that the interfacial thermal conductance is comparable, or even higher than that between graphene and common metals. Subsequently, the effects of temperature and vacancy defects on the interfacial thermal conductance are also investigated. Our findings suggest that the designed MoS$_2$-graphene in-plane heterostructure is promising to overcome the bottleneck in the thermal management of MoS$_2$-based integrated devices.

## COMPUTATIONAL METHODS

**DFT calculations of atomic interfacial structure**

The density functional theory (DFT) calculations were performed by using the Vienna Ab-initio Simulation Package (VASP)[21,22] to relax the atomic structures and obtain the binding energy of MoS$_2$-graphene contact. Generalized gradient approximation (GGA) with the Perdew–Burke–Ernzerhof (PBE) functional[23] was used to describe the exchange-correlation interaction. Projector-augmented wave (PAW) technology[24] was used to describe the core electrons. A plane-wave basis kinetic energy cutoff of 400 eV and a convergence criterion of $10^{-4}$ eV were used in the calculations. For the DFT calculations, both graphene and MoS$_2$ ribbons include six ZZ rows in their width. Hydrogen atoms were used to passivate both the ZZ edge of graphene and ZZ S edge of MoS$_2$ in opposite to the contacting junction. All atoms were fully relaxed until the force is lower than 0.02 eV/Å.

**MD calculations of interfacial thermal transport**

MD simulations are employed to study the thermal transport across the MoS$_2$-graphene heterostructures using the LAMMPS package[25]. In all the MD simulations performed here, we used the Adaptive Intermolecular Reactive Empirical Bond Order (AIREBO) interatomic potential[26] to describe the reactive, covalent bonding interactions within MoS$_2$ and graphene, respectively. This potential was parameterized to match the DFT calculations of Mo-S interaction in MoS$_2$[27,28], and C-C interaction in graphene[29], and has been widely used to study the structural, mechanical and thermal properties of MoS$_2$[28,30–32] and graphene[29,33,34]. The atomic interaction between Mo-C was described by Morse potential[35]. The velocity Verlet algorithm was employed to integrate Newton's equations of atom motion, and the MD time step was set as 0.5 fs. The system was optimized at 300 K for 100 ps to achieve the contact structure.

In MD simulation, the total heat flux $J$ in the longitudinal direction was obtained by[36]

$$J = \frac{1}{V}\left[\sum_i^N \varepsilon_i \mathbf{v}_i + \frac{1}{2}\sum_{ij;i\neq j}^N (\mathbf{F}_{ij}\cdot \mathbf{v}_i)\mathbf{r}_{ij} + \frac{1}{6}\sum_{ijk;i\neq j\neq k}^N (\mathbf{F}_{ijk}\cdot \mathbf{v}_i)(\mathbf{r}_{ij}+\mathbf{r}_{ik})\right], \quad (1)$$

where, $\varepsilon_i$ and $\mathbf{v}_i$ are the energy density and velocity associated with atom $i$, respectively. Vector $\mathbf{r}_{ij}$ denotes the interatomic distance between atoms $i$ and $j$, and $\mathbf{F}_{ij}$ and $\mathbf{F}_{ijk}$ denote the two-body and three-body force, respectively. $V$ is the volume of the studied system. After the system reached the non-equilibrium steady state, a time averaging of temperature and heat flux was performed for an additional 20 ns. Note that we excluded the regions near two heat reservoirs and computed the heat flux only for the rest of the system to avoid the boundary effect from the heat source and sink.

## RESULTS AND DISCUSSION

### In-plane interfacial binding between $MoS_2$ and graphene

The schematic of the interface construction is shown in Figure 1a. In general, there are two major edge types for graphene and $MoS_2$: Zigzag and armchair. Since previous theoretical and experimental studies have shown that zigzag (ZZ) edges possess a lower formation energy and are more stable than armchair edges for both graphene[37,38] and $MoS_2$[39,40] in the growth, here only the zigzag-oriented interfaces are considered, which are named as ZZ interface, as shown in Figure 1. Note that in addition to the interfaces formed by pristine ZZ edges as shown in Figure1b and 1c, the (5|7) reconstructed zigzag edge of graphene is also considered in accordance to experimental observation[41], and its interfaces with $MoS_2$ are named as ZZ57 interface, as shown in Figure 1d and 1e.

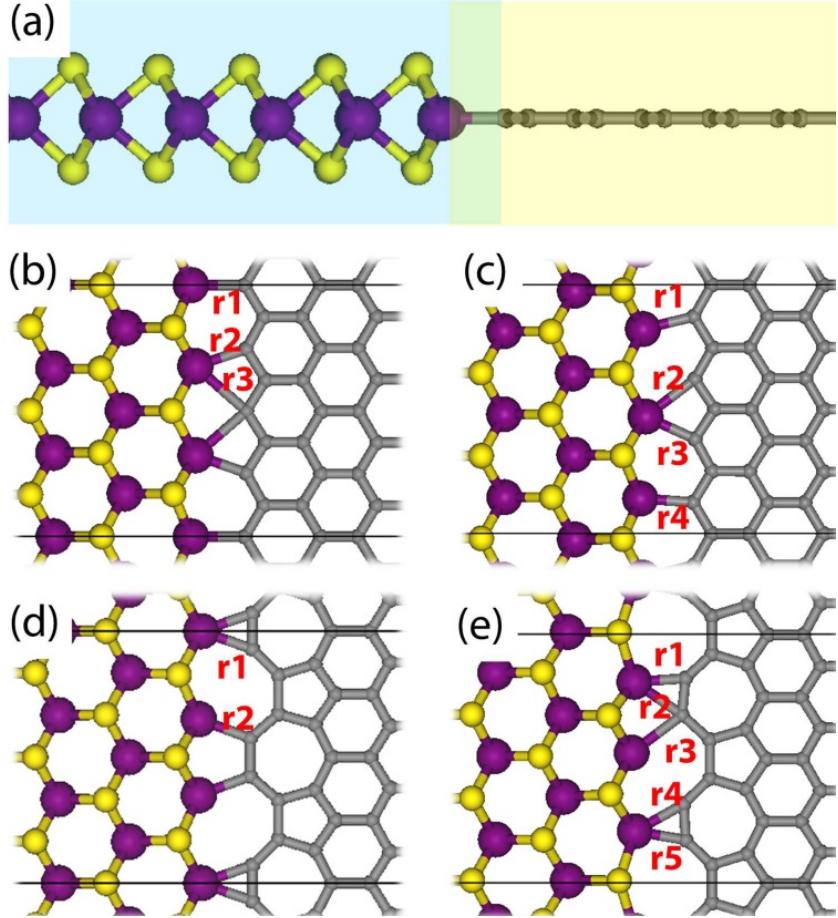

**Figure 1. Interface structures of in-plane graphene-MoS$_2$ heterostructures.** The side view (**a**) of the interface between the MoS$_2$ and graphene edge, and the top view of four probable interface configurations, ZZ-1 (**b**), ZZ-2(**c**), ZZ57-1(**d**), ZZ57-2 (**e**). Mo, S, C atoms are shown in purple, yellow, and gray color, respectively.

The lattice constants of graphene and MoS$_2$ are $a_G$ = 2.46 Å and $a_{MS}$ = 3.20 Å, respectively. Thus, along the interface, $4a_G$ graphene is interfacing with $3a_{MS}$ MoS$_2$ with a small lattice mismatch strain of 2.4%. To find the lowest-energy interface configuration, the graphene edge is allowed to move relatively with respect to the MoS$_2$ edge, generating a series of initial interface structures. After energy relaxation, for both ZZ and ZZ57 interfaces, the initial structures fall into two typical configurations, which are shown in Figure 1b and 1c, and Figure 1d and 1e, respectively. For all these four configurations, no significant ripple/buckling was observed near the interfaces. The binding energies between graphene ZZ (ZZ57) and MoS$_2$ edges are about 12.27 eV/nm (8.87 eV/nm), i.e. 3.953 eV/Mo (2.858 eV/Mo), indicating a very strong interaction between graphene and $MoS_2$. Such strong interaction between Mo and C atoms is found to be covalent bonding in nature, suggesting that the in-plane $MoS_2$-graphene interface is energetically stable.

**Table 1. The binding energy and bond lengths (see Figure 1) between $MoS_2$ edge and graphene edge.** The binding energy is calculated as $E_b = (E_G + E_{MoS_2} - E_{Tot})/L$ or $E_b' = (E_G + E_{MoS_2} - E_{Tot})/N_{Mo}$, where $E_G$, $E_{MoS2}$ and $E_{Tot}$ are the energy of graphene ribbon with ZZ or ZZ57 edge, $MoS_2$ ribbon and the whole system, respectively. $L$ is the length along the $MoS_2$-graphene, and $N_{Mo}$ is the number of edge Mo atoms.

|         | $E_b$ (eV/nm) | $E_b'$ (eV/Mo) | r1 (Å) | r2 (Å) | r3 (Å) | r4 (Å) | r5 (Å) |
|---------|---------------|----------------|--------|--------|--------|--------|--------|
| **ZZ-1**   | 12.27 | 3.953 | 2.073 | 2.062 | 2.599 | –     | –     |
| **ZZ-2**   | 12.25 | 3.948 | 2.039 | 2.405 | 2.111 | 2.065 | –     |
| **ZZ57-1** | 8.87  | 2.858 | 2.075 | 2.100 | –     | –     | –     |
| **ZZ57-2** | 8.37  | 2.697 | 1.987 | 2.257 | 2.445 | 2.166 | 2.004 |

To explore the stability of these two types of graphene-$MoS_2$ interface, we compared the energy difference between the lowest-energy ZZ and ZZ57 interface: $\Delta E = E_{Tot}(ZZ) - E_{Tot}(ZZ57) = -0.9$ eV/nm, i.e. the interface energy of the ZZ interface is 0.9 eV/nm lower than that of the ZZ57 interface, indicating that the (5|7) configuration is not the most energetically favorable. This can be explained by the fact that the dangling bonds of graphene ZZ edge can be saturated by the Mo edge, similar to the phenomenon reported when graphene ZZ edge interacts with transition metal surfaces[42]. The detailed bonding situations obtained from the DFT calculations are given in Figure 1b-1e and also Table 1, which are then used to build large models of $MoS_2$-graphene in-plane heterostructure for molecular dynamics simulations.

**Interfacial thermal conductance\**

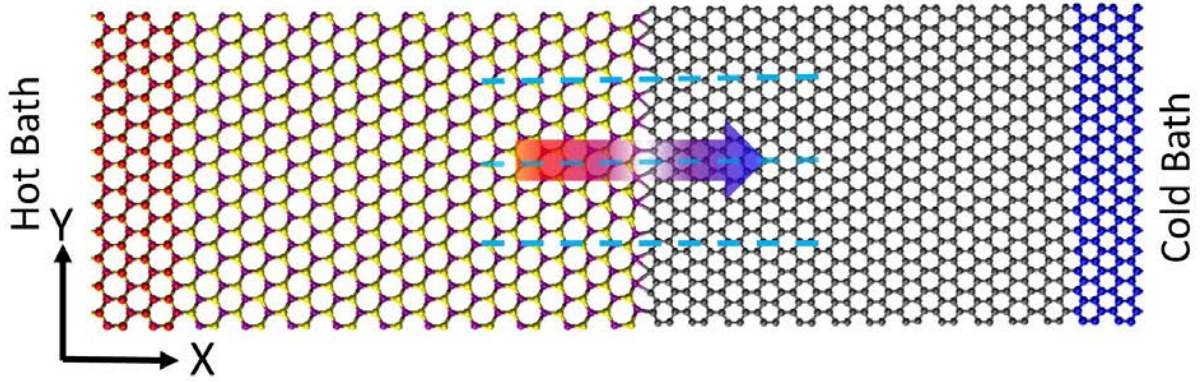

**Figure 2. Schematic of MoS$_2$-graphene heterostructure with hot and cold reservoirs at the two ends.**

We computed the thermal conductance of the system using non-equilibrium molecular dynamics (NEMD) simulations. Schematic of the atomistic model for the MoS$_2$-graphene in-plane heterostructure used in the NEMD simulation is shown in Figure 2. According to our DFT calculations, ZZ-1 and ZZ57-1 interfacial configurations have relatively higher binding energies, thus, here we only consider these two interfacial configurations. After the relaxation of ZZ-1 configuration, we find the binding energy to be about 3.778 eV/Mo, which is in good agreement with our first-principles calculation result (3.953 eV/Mo). In all the simulated heterostructures, the lengths of graphene and MoS$_2$ are $L_G$ = 106.52 Å and $L_{MS}$ = 109.12 Å, respectively. The width of the heterostructures is W = 38.61 Å, which is four times of the width of the structure in our DFT calculations.

Here we set the *x* axis to be the direction of the heat current, and the *y* axis to be parallel to the MoS$_2$-graphene interfaces. To eliminate the edge effects on thermal transport, periodic boundary condition is applied along the *y* direction. To establish a temperature gradient along the longitudinal *x* direction, the atoms close to the left end (of MoS$_2$) and the right end (of graphene) were placed into hot and cold Nosé-Hoover reservoirs[43] with temperatures set to be $T_H$ = 310 K and $T_C$ = 290 K, respectively, as shown in Figure 2. The simulations were then performed long enough (1.25 ns) to allow the system to reach non-equilibrium steady state, where the temperature gradient was well established, and the heat flux going through the system became time-independent.

A typical temperature profile at steady state in the MoS$_2$-graphene heterostructure at 300 K is shown in Figure 3. The temperature profile represents the heat energy transport at different sections of the heterostructure. At non-equilibrium steady state, the distributions of temperature and heat flux are time-independent. As the width of graphene section is the same as that of the MoS$_2$ section, both total heat current and heat flux density are the same in these two sections. Thus the temperature gradient is directly related to the thermal conductivity: The large temperature gradient in MoS$_2$ part is the result of its low thermal conductivity; and the small temperature gradient in graphene is originated from its ultra-high thermal conductivity. Due to the large difference in lattice (thermal) properties between graphene and MoS$_2$, there is a remarkable temperature jump $\delta T$ across the interface, indicating the existence of interfacial thermal resistance. In NEMD, the value of interfacial thermal conductance (ITC) can be calculated by $\lambda = J/\delta T$, where $\lambda$ is the ITC, and $J$ is the heat flux across the interface. The ITC of the ZZ-1 interface is 2.49×10$^8$ WK$^{-1}$m$^{-2}$ at room temperature. This value is comparable with that of chemically-bonded graphene-metal interfaces (2.5×10$^8$ WK$^{-1}$m$^{-2}$)[44] predicted by first-principles calculations, indicating that MoS$_2$-graphene in-plane heterostructure is efficient for heat transport. The interfacial thermal conductance between graphene and MoS$_2$ in the vertical heterostructure via van der Waals interaction has been calculated by MD simulations[34], and the ITC was only 0.138×10$^8$ WK$^{-1}$m$^{-2}$. Promisingly, the high ITC in the MoS$_2$-graphene in-plane heterostructure (one order higher than the interlayer thermal conductance), together with its high structural stability, may provide a viable solution for thermal management in MoS$_2$-based electronic devices. Our calculation shows that the ITC of the ZZ57 interface is 2.22×10$^8$ WK$^{-1}$m$^{-2}$ at room temperature, which is about 12% lower than that of the ZZ interface.

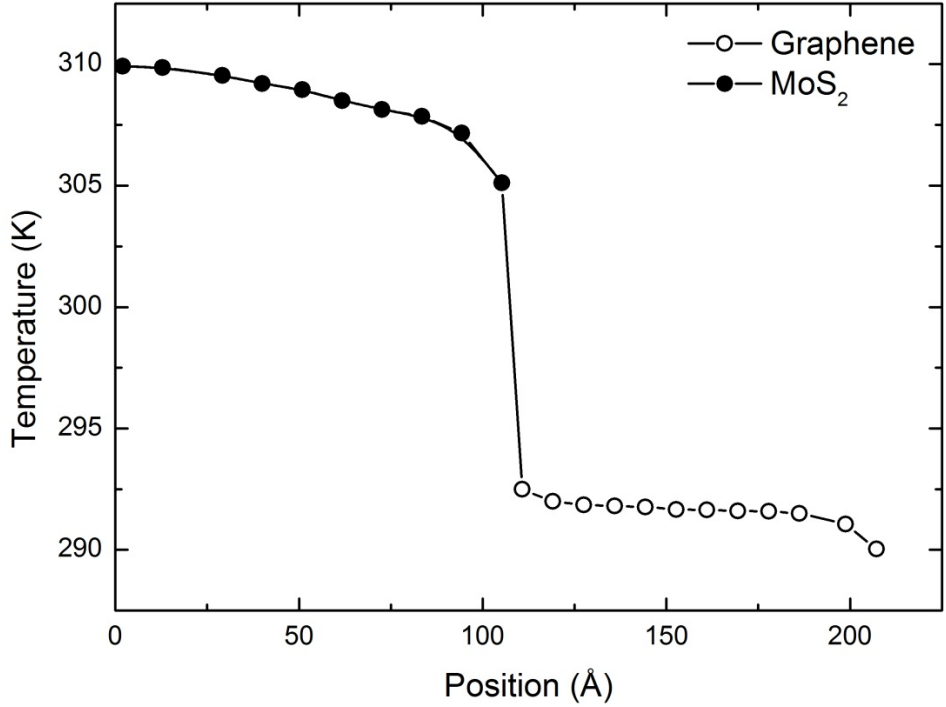

**Figure 3. Temperature profile in the heat flux direction from MoS$_2$ to graphene.** The solid and hollow dots represent the temperature profiles in MoS$_2$ and graphene, respectively.

To understand the difference of interfacial thermal conductance at ZZ and ZZ57 interfaces, the phonon transmission process across the interfaces was investigated using the phonon wave packet method[45,46]. Phonon wave packets are formed from a linear combination of vibration eigenstates of the crystalline lattices. A wave packet centered at $k_0$ in $k$ space and $x_0$ in real space is generated by setting the atom displacement as:

$$u_n = A\varepsilon(k_0)\exp[i k_0(x_n - x_0)]\exp[-(x_n-x_0)^2/\xi^2] \quad (2)$$

where $u_n$ is the displacement of the $n$th atom, $A$ is amplitude of the wave, $\varepsilon$ is the polarization vector, $\xi$ (=100$a$) is the width of the wave packet. The key idea of this method is to construct a phonon wave packet from a single branch of the phonon dispersion curve with a narrow frequency range and well-defined polarization.

Upon encounter with an interface, the wave packet is scattered into transmitted and reflected waves. By computing the ratio of the amplitudes of transmitted ($A_{tr}$) over initial ($A$) phonon waves at the interface, the energy transmission coefficient $\alpha$ can be determined by:

$$\alpha = \left(\frac{A_{tr}}{A}\right)^2 \qquad (3)$$

In our realization, a 705.1 nm long sample is used. Here we generate a wave packet in graphene by disturbing the atoms according to Eq. (2), and then record the average atomic displacement in each atom as the system evolves with time.

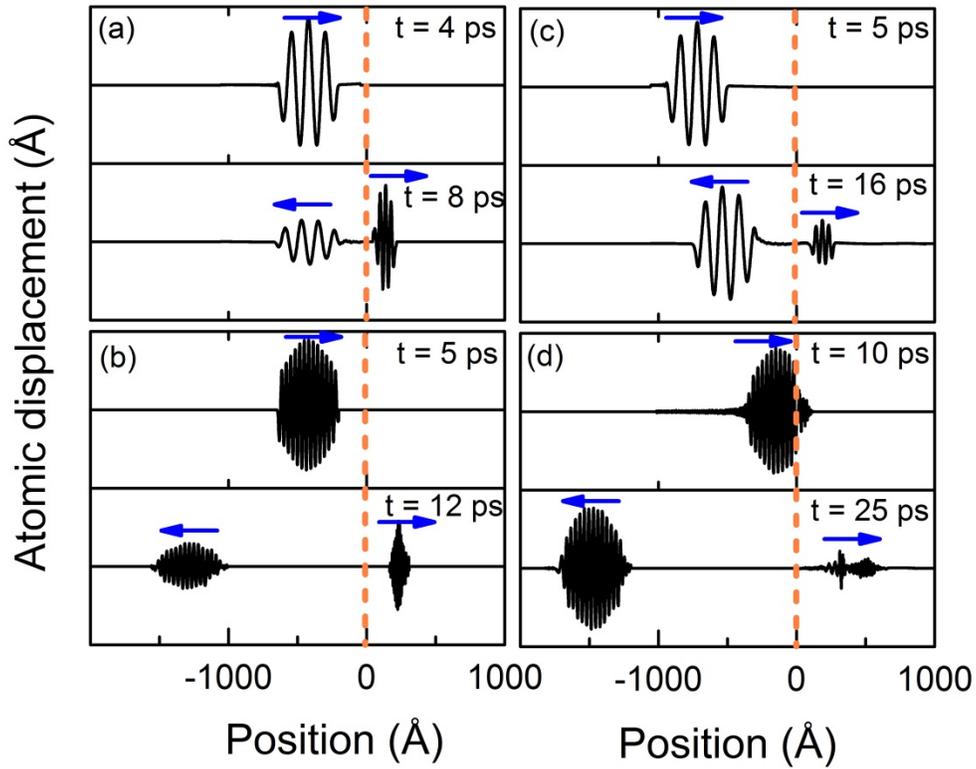

**Figure 4. Snapshots of a wave packet across the ZZ interface.** The blue arrows indicate the travelling direction of the wave packet, and brown dashed lines denote the interface. (a) LA, $k_0 = 0.02 \times 2\pi/a$, $f = 1.90$THz. (b) LA, $k_0 = 0.075 \times 2\pi/a$, $f = 7.10$THz. (c) TA, $k_0 = 0.02 \times 2\pi/a$, $f = 1.18$THz. (d) LA, $k_0 = 0.075 \times 2\pi/a$, $f = 4.41$THz.

Two dominant acoustic phonon modes (LA and TA) are studied here, and the snapshots of a wave packet near the ZZ interfaces are given in Figure 4. With the same wave vector $k_0 = 0.02(\times 2\pi/a)$, about 75% of phonon energy from LA phonon wave packet transmits across the interface, while only about 12% of energy from TA phonon wave packet transmits across the interface. Figure 5 shows the transmission coefficients for phonons with different frequencies. The inset of Figure 5 shows the phonon dispersions of graphene and $MoS_2$. It is seen that the interface effectively blocks high frequency phonon modes due to the lack of LA and TA energy states in $MoS_2$ at the same frequencies. For LA phonon mode, when the incident frequencies are lower than 7.3 THz, there is no remarkable change in phonon transmission coefficients; however, when it reaches 8 THz, the phonon transmission dramatically decreases. A total reflection occurs when the frequency is higher than 9 THz. In contrast to LA phonon mode, even in the low frequency regime, the transmission coefficient of TA phonons is substantially lower than that of LA phonons, which indicates that LA modes are the dominant contribution to the heat transport across the interface.

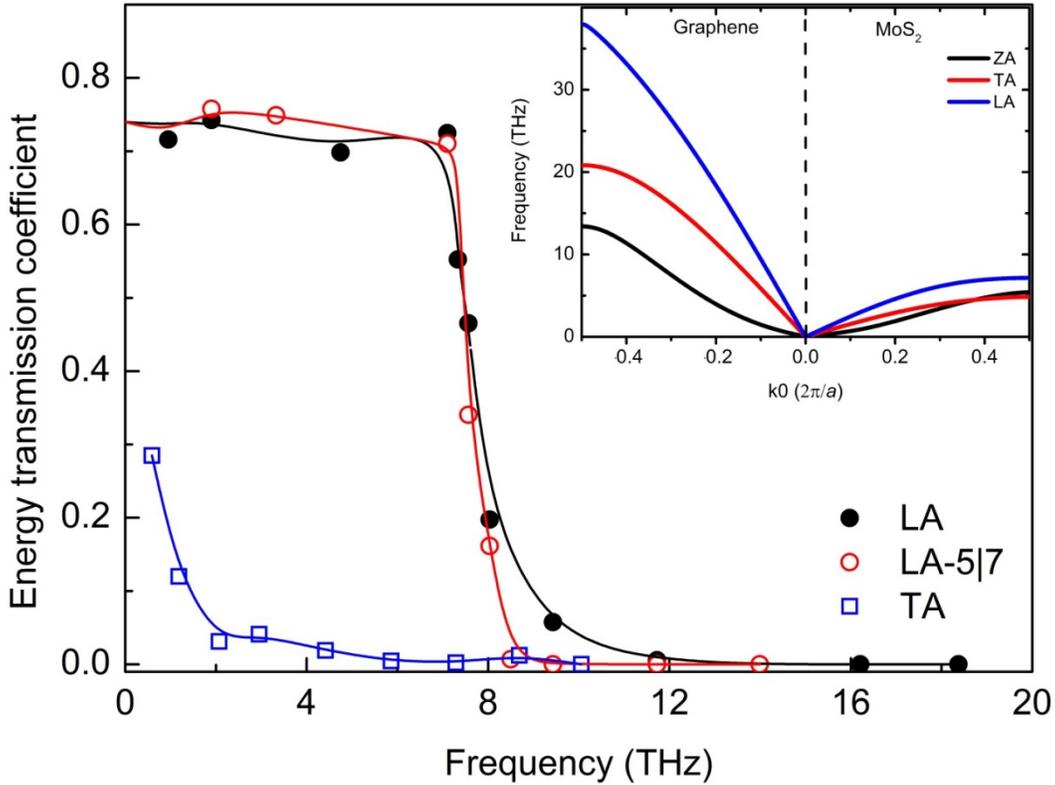

**Figure 5. Energy transmission coefficient α of a phonon wave packet as a function of frequency.** The inset displays the phonon dispersion of graphene and $MoS_2$. The $MoS_2$ phonon dispersion is adopted from Ref. [47] by first principle calculations.

Moreover, at the ZZ57 interface, for the low frequency LA phonons (f < 7.3THz), there is no significant difference in the transmission coefficient α (~0.75) with respect to the ZZ interfaces. However, in the high frequency regime, the transmission coefficient of ZZ57 interface decreases quickly. For example, with the same frequency $f$ = 7.6THz, $α_{zz57}$ is only 0.34, which is 28% lower than $α_{zz}$ (~0.47). This difference is likely due to the difference in interface atomic structures between them. For the ZZ57 interface, besides phonon scattering at the Mo-C interface, there exists extra phonon scattering from the 5|7 defects at the interface, which certainly reduces the phonon transport.

**Effect of vacancy on interfacial thermal transport**

Atomic vacancies are often present at the interfaces of two-dimensional heterostructures.[48,49] It is well-known that vacancies are able to disrupt regular atomic structures and cause additional phonon scattering[50]. Therefore, here, we focus on Mo vacancy at the interface to understand the effect of Mo vacancy concentration on the thermal transport across the $MoS_2$-graphene interfaces at room temperature.

Figure 6a shows the interfacial thermal conductance as a function of Mo vacancy concentration at the ZZ and ZZ57 interfaces. Clearly, introduction of Mo vacancies at the interface leads to a decrease in the density of Mo-C covalent bonds. As a result, the ITC decreases linearly with the Mo vacancy concentration. From the slopes of the linear dependence, we find that each covalent bond serves as an independent channel for the heat transport, with a constant thermal conductance of $1.48 \times 10^7$ $WK^{-1}m^{-2}$ and $1.38 \times 10^7$ $WK^{-1}m^{-2}$ for the ZZ and ZZ57 interfaces, respectively. The linear dependence of ITC on the Mo vacancy concentration provides an effective route to modulate the heat transport across the interface.

As shown in Figure 6b, the temperature jump increases linearly with increasing the Mo vacancy concentration. With the same Mo vacancy concentrations, the ZZ and ZZ57 interfaces have the same temperature jump; however, the heat flux across the ZZ interface is higher than that cross the ZZ57 interface. As a result, the ITC of the ZZ interface is higher.

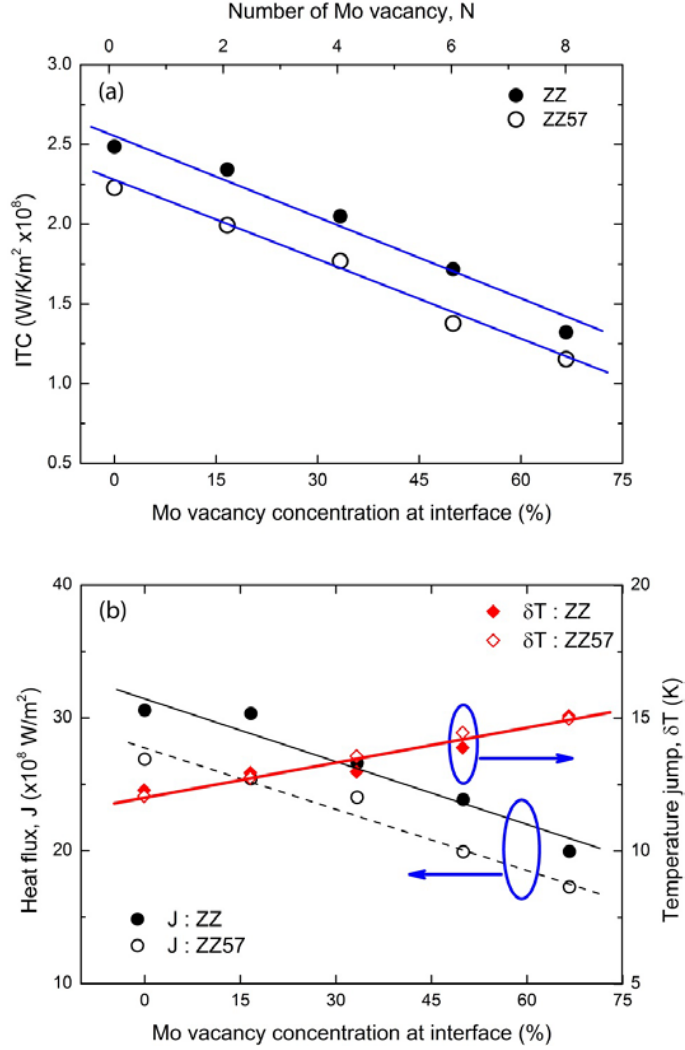

**Figure 6**. **Mo vacancy-dependent interfacial thermal conductance, heat flux, and temperature jump.** (**a**) Interfacial thermal conductance as a function of Mo vacancy concentration at the interface. (**b**) Mo vacancy concentration-dependent interfacial temperature jump and heat flux.

**Effect of temperature on interfacial thermal transport**

Next, we explore the temperature effect on the ITC for the ZZ and ZZ57 interfaces. As shown in Figure 7, it is clear that values of ITC at both ZZ and ZZ57 interfaces increase with temperature. This is due to the fact that with increasing temperature, more high-frequency phonons are excited, providing additional carriers for the interfacial thermal transport. Moreover, the anharmonicity of atomic vibrations at the interface also increases with increasing temperature. As a result, the phonon transmission coefficient is enhanced through inelastic phonon scattering[51]. Hence, both mechanisms favor the increase in ITC.

It is interesting to point out that the ITCs of the ZZ and ZZ57 interfaces exhibit different increasing trends with temperature. When the temperature is below 400 K, the ITC profile of the ZZ interface possesses a similar slope as that of the ZZ57 interface. However, when the temperature is higher than 400 K, the ITC of the ZZ interface increases superlinearly, while the ITC of the ZZ57 interface increases sublinearly. As temperature increases from 400K to 500K, the ITC of the ZZ increases by 33%, from $3.20 \times 10^8$ $WK^{-1}m^{-2}$ to about $4.25 \times 10^8$ $WK^{-1}m^{-2}$. However, in the same temperature range, the ITC of the ZZ57 increases only 7%, from about $2.84 \times 10^8$ $WK^{-1}m^{-2}$ to $3.05 \times 10^8$ $WK^{-1}m^{-2}$. This phenomenon can be understood from the effect of atomic defects on phonon scattering, and the dependence on phonon wavelength. It is known that the long-wavelength (low frequency) phonons are insensitive to atomic scale defects. As a result, the atomic scale defects have negligible influence on the temperature-dependence of ITC at low temperature range. However, at high temperature range, more short-wavelength (high frequency) phonons are excited, which are more sensitive to the atomic defects. Hence, at the ZZ57 interface, as the temperature increases, the increased distribution from high-frequency phonons is over-run by the enhanced phonon scattering at 5|7 defects. Therefore, the ITC of the ZZ57 interface becomes sublinear with increasing temperature.

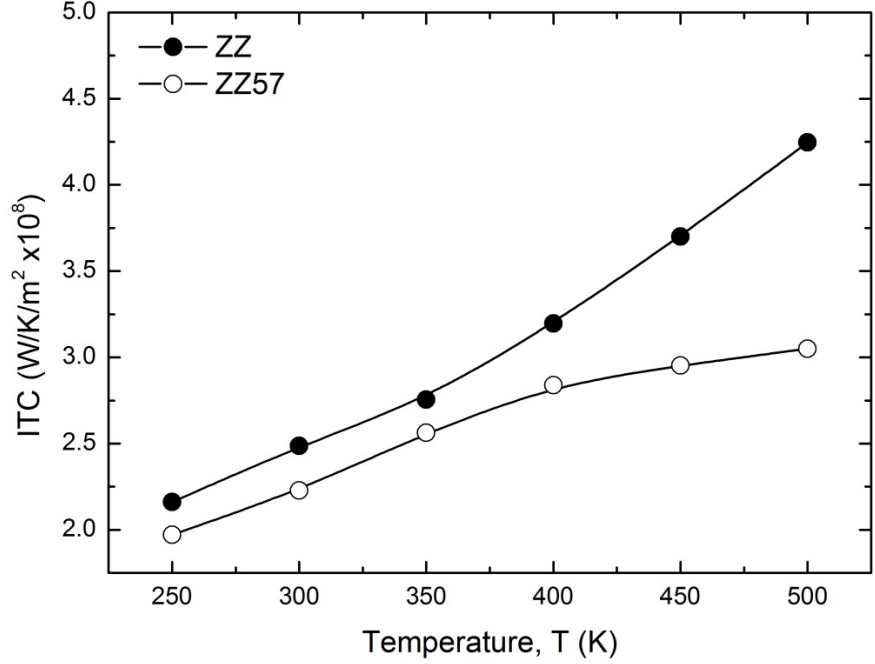

**Figure 7. Interfacial thermal conductance as a function of temperature.**

In addition to the interfacial configuration, the structural size may also have a remarkable effect on the interfacial thermal conductance of nanoscale heterostructures[52]. To understand this effect, we perform simulations by increasing the length of graphene section from 106.52Å to 213.04 Å. It is found that the interfacial thermal conductance increases with length of graphene, from $2.49 \times 10^8$ WK$^{-1}$m$^{-2}$ (106.52Å) to $2.68 \times 10^8$ WK$^{-1}$m$^{-2}$ (159.78 Å) further to $2.69 \times 10^8$ WK$^{-1}$m$^{-2}$ (213.04 Å), respectively. This trend is consistent with many previous works on the length effect on the interfacial thermal conductance, which arises from more phonon modes available for thermal transport across the interface in the longer samples[53–56]. Similarly, the total thermal conductance also increases due to more phonon modes available for thermal transport in the longer samples[57–59]. Therefore, a longer graphene section is able to promote the total heat transport and enhance the thermal management efficiency of MoS$_2$ devices.

## CONCLUSION

We have designed 2D heterostructures by constructing the interfaces between semiconducting MoS$_2$ and conducting graphene. First-principles calculations have been performed to examine the bonding features and energetic stability of these heterostructures. The calculated binding energies suggest that strong covalent bonds are formed between Mo and C atoms at MoS$_2$-graphene interfaces. Using the atomistic structures predicted by DFT calculation, we have investigated the interfacial thermal conductance of these heterostructures using molecular dynamics simulations. A remarkably high interfacial thermal conductance is found, which is beneficial to the thermal management of MoS$_2$ integrated devices. By introducing Mo vacancies at the interfaces, we find that the ITC decreases linearly with increasing the Mo vacancy concentration. This provides an effective route to control the phonon transport channels and ITC. In addition, the ITC is found to increase with the temperature, which can be attributed to the enhanced inelastic phonon scattering and more excited phonons at higher temperature. Our work not only presents stable MoS$_2$-graphene in-plane heterostructures, but also proposes a viable solution to overcome the bottleneck in thermal management of MoS$_2$ integrated devices.

**Acknowledgments:** We gratefully acknowledge the financial support from the Agency for Science, Technology and Research (A*STAR), Singapore and the use of computing resources at the A*STAR Computational Resource Centre, Singapore. This work was supported in part by a grant from the Science and Engineering Research Council (152-70-00017).

**Competing interests:** The authors declare that they have no competing interests.

# REFERENCES

1. Geim, A. K.; Grigorieva, I. V. *Nature.* **2013**, *499***,** 419−425.
2. Zhao J., Liu H., Yu, Zh., Quhe, R., Zhou, S., Wang, Y., Liu, C. C., Zhong, Ho., Han, N., Lu, J., Yao, Y., Wu, K. *Prog. Mater. Sci.* **2016**, *83***,** 24−151.
3. Allain, A.; Kang, J.; Banerjee, K.; Kis, A. *Nat. Mater.* **2015**, *14***,** 1195−1205.
4. Kang, K.; Xie, S.; Huang, L.; Han, Y.; Huang, P. Y.; Mak, K. F.; Kim, C.-J.; Muller, D.; Park, J. *Nature* **2015**, *520***,** 656−660.
5. Yu, Z.; Pan, Y.; Shen, Y.; Wang, Z.; Ong, Z.-Y.; Xu, T.; Xin, R.; Pan, L.; Wang, B.; Sun, L.; Wang, J.; Zhang, G.; Zhang, Y.-W.; Shi, Y.; Wang, X. *Nat. Commun.* **2014**, *5***,** 5290.
6. Cui, X.; Lee, G.-H.; Kim, Y. D.; Arefe, G.; Huang, P. Y.; Lee, C.-H.; Chenet, D. A.; Zhang, X.; Wang, L.; Ye, F.; Pizzocchero, F.; Jessen, B. S.; Watanabe, K.; Taniguchi, T.; Muller, D. A.; Low, T.; Kim, P.; Hone, J. *Nat. Nano.* **2015**, *10***,** 534−540.
7. Liu, X.; Wu, H.; Cheng, H.-C.; Yang, S.; Zhu, E.; He, Q.; Ding, M.; Li, D.; Guo, J.; Weiss, N. O.; Huang, Y.; Duan, X. *Nano Lett.* **2015**, *15***,** 3030−3034.
8. Yu, Z.; Ong, Z.-Y.; Pan, Y.; Cui, Y.; Xin, R.; Shi, Y.; Wang, B.; Wu, Y.; Chen, T.; Zhang, Y.-W.; Zhang, G.; Wang, X. *Adv. Mater.* **2016***, 28***,** 547−552.
9. Sahoo, S.; Gaur, A. P. S.; Ahmadi, M.; Guinel, M. J. F.; Katiyar, R. S. *J. Phys. Chem. C* **2013**, *117***,** 9042−9047.
10. Yan, R.; Simpson, J. R.; Bertolazzi, S.; Brivio, J.; Watson, M.; Wu, X.; Kis, A.; Luo, T.; Walker, A. R. H.; Xiang, H. G. *ACS Nano* **2014**, *8***,** 986−993.
11. Taube, A.; Judek, J.; Łapińska, A.; Zdrojek, M. *ACS Appl. Mater. Interfaces* **2015**, *7***,** 5061−5065.
12. Liu, X.; Zhang, G.; Pei, Q.-X.; Zhang, Y.-W. *Appl. Phys. Lett.* **2013**, *103***,** 133113.
13. Li, W.; Carrete, J.; Mingo, N. *Appl. Phys. Lett.* **2013***, 103***,** 253103.
14. Cai, Y.; Lan, J.; Zhang, G.; Zhang, Y.-W. *Phys. Rev. B* **2014**, *89,* 035438.
15. Jiang, J.-W.; Zhuang, X.; Rabczuk, T. *Sci. Rep.* **2013**, *3,* 2209.
16. Wu, X.; Yang, N.; Luo, T. *Appl. Phys. Lett.* **2015***, 107,* 191907.
17. Balandin, A. A.; Ghosh, S.; Bao, W.; Calizo, I.; Teweldebrhan, D.; Miao, F.; Lau, C. N. *Nano Lett.* **2008**, *8,* 902.
18. Ghosh, S.; Bao, W.; Nika, D. L.; Subrina, S.; Pokatilov, E. P.; Lau, N.; Balandin, A. A. *Nat. Mater.* **2010**, *9,* 555−558.

19. Seol, J. H.; Jo, I.; Moore, A. L.; Lindsay, L.; Aitken, Z. H.; Pettes, M. T.; Li, X.; Yao, Z.; Huang, R.; Broido, D.; Mingo, N.; Ruoff, R. S.; Shi, L. *Science* **2010**, *328,* 213.
20. Sadeghi, M. M.; Jo, I.; Shi, L. *Proc. Natl. Acad. Sci. U.S.A.* **2013**, *110,* 16321−16326.
21. Kresse, G.; Furthmüller, J. *Phys. Rev. B* **1996**, *54*, 11169.
22. Kresse, G.; Furthmüller, J. *Comp. Mater. Sci.* **1996**, *6,* 15−50.
23. Perdew, J. P.; Burke, K.; Ernzerhof, M. *Phys. Rev. Lett.* **1996**, *77,* 3865.
24. Blöchi, P. E. *Phys. Rev. B* **1994**, *50,* 17953.
25. Plimpton, S. *J. Comput. Phys.* **1995**, *117,* 1−19.
26. Stuart, S. J.; Tutein, A. B.; Harrison, J. A. *J. Chem. Phys.* **2000**, *112,* 6472-6486.
27. Liang, T.; Phillpot, S. R.; Sinnott, S. B. *Phys. Rev. B* **2009**, *79,* 245110.
28. Stewart, J. A.; Spearot, D. E. *Modelling Simul. Mater. Sci. Eng.* **2013**, *21,* 045003.
29. Liu, X.; Zhang, G.; Zhang, Y.-W. *J. Phys. Chem. C* **2014**, *118,* 12541−12547.
30. Tang, D.-M.; Kvashnin, D. G.; Najmaei, S.; Bando, Y.; Kimoto, K.; Koskinen, P.; Ajayan, P. M.; Yakobson, B. I.; Sorokin, P. B.; Lou, J.; Golberg, D. *Nat. Commun.* **2014**, *5,* 3631.
31. Dang, K. Q.; Spearot, D. E. *J. Appl. Phys.* **2014**, *116,* 013508.
32. Wang, X.; Tabarraei, A.; Spearot, D. E. *Nanotechnology* **2015**, *26,* 175703.
33. Liu, X.; Zhang, G.; Zhang, Y.-W. *Nano Res.* **2015**, *8,* 2755−2762.
34. Ding, Z.; Pei, Q.-X.; Jiang, J.-W.; Huang, W.; Zhang, Y.-W. *Carbon* **2016**, *96,* 888−896.
35. Rappé, A. K.; Casewit, C. J.; Colwell, K. S.; Goddard III, W. A.; Skiff, W. M. *J. Am. Chem. Soc.* **1992**, *114,* 10024−10035.
36. Volz, S. G.; Chen, G. *Appl. Phys. Lett.* **1999**, *75,* 2056-2058.
37. Gao, Y.; Zhang, Y.; Chen, P.; Li, Y.; Liu, M.; Gao, T.; Ma, D.; Chen, Y.; Cheng, Z.; Qiu, X.; Duan, W.; Liu, Z. *Nano Lett.* **2013**, *13,* 3439−3443.
38. Liu, M.; Li, Y.; Chen, P.; Sun, J.; Ma, D.; Li, Q.; Gao, T.; Gao, Y.; Cheng, Z.; Qiu, X. Fang, Y.; Zhang, Y.; Liu, Z. *Nano Lett.* **2014**, *14,* 6342−6347.
39. Li, Y.; Zhou, Z.; Zhang, S.; Chen, Z. *J. Am. Chem. Soc.* **2008**, *130,* 16739−16744.
40. Wang, Z.; Li, H.; Liu, Z.; Shi, Z.; Lu, J.; Suenaga, K.; Joung, S.-K.; Okazaki, T.; Gu, Z.; Zhou, J.; Gao, Z.; Li, G.; Sanvito, S.; Wang, E.; Iijima, S. *J. Am. Chem. Soc.* **2010**, *132,* 13840−13847.
41. Girit, Ç.; Meyer, J. C.; Erni, R.; Rossell, M. D.; Kisielowski, C.; Yang, L. C.-H.; Park, M. F.; Cohen, M. L.; Louie, S. G.; Zettl, A. *Science* **2009**, *323,* 1705−1708.

42. Gao, J.; Zhao, J.; Ding, F. *J. Am. Chem. Soc*. **2012**, *134,* 6204−6209.
43. Nosé, S. *J. Chem. Phys*. **1984**, *81,* 511−519.
44. Mao, R.; Kong, B. D.; Gong, C.; Xu, S.; Jayasekera, T.; Cho, K.; Kim, K. W. *Phys. Rev. B*, **2013**, *87,* 165410.
45. Schelling, P. K.; Phillpot, S. R.; Keblinski, P. *Appl. Phys. Lett*. **2002**, *80*, 2484−2486.
46. Xu, W.; Zhang, G.; Li, B. *J. App. Phys*. **2014**, *116*, 134303.
47. Cai, Y.; Lan, J.; Zhang, G.; Zhang, Y.-W. *Phys. Rev. B* **2014**, *89*, 035438.
48. Drost, R.; Kezilebieke, S.; Ervasti, M. M.; Hämäläinen, S. K.; Schulz, F.; Harju, A.; Liljeroth, P. *Sci. Rep*. **2015**, *5,* 16741.
49. Feng, L.; Su, J.; Liu, Z.-T. *RSC Adv*. **2015**, *5,* 20538−20544.
50. Xie, G.; Shen, Y.; Wei, X.; Yang, L.; Xiao, H.; Zhong, J.; Zhang, G. *Sci. Rep.* **2014**, *4,* 5085.
51. Hu, L.; Desai, T.; Keblinski, P. *Phys. Rev. B*, **2011**, *83,* 195423 (2011).
52. Duan, X.; Wang, C.; Shaw, J. C.; Cheng, R.; Chen, Y.; Li, H.; Wu, X.; Tang, Y.; Zhang, Q.; Pan, A.; Jiang, J.; Yu, R.; Huang, Y.; Duan, X. *Nat. Nano*. **2014**, *9*, 1024−1030.
53. Bagri, A.; Kim, S. P.; Ruoff, R. S.; Shenoy, V. B. *Nano Lett*. **2010**, *11*, 3917.
54. Chen, J.; Zhang, G.; Li, B. *J. Appl. Phys*. **2012**, *112*, 064319.
55. Jones, R. E.; Duda, J. C.; Zhou, X. W.; Kimmer, C. J.; Hopkins, P. E. *Appl. Phys. Lett*. **2013**, *102*, 183119.
56. Liu, X.; Zhang, G.; Zhang, Y.-W. *Nano Research*, **2016**, *9*, 2372–2383.
57. Xu, Y; Chen, X; Gu, B.-L.; Duan, W.-H. *Appl. Phys. Lett*., 2009, 95(23): 233116.
58. Nika, D. L.; Askerov, A. S.; Balandin, A. A. *Nano Lett*. **2012**, *12*, 3238.
59. Yu, C.; Zhang, G.; *J. Appl. Phys*. **2013**, *113*, 044306.